\documentclass{iau}
\usepackage{graphicx}
\usepackage{hyperref}
\usepackage{natbib}

\def\apj{ApJ}
\def\aj{AJ}
\def\mnras{Mon. Not. R. Astron. Soc.}

\def\apjs{Astrophys. J. Supp.}

\title[Bayesian large-scale structure inference: initial conditions and the cosmic web] 
{Bayesian large-scale structure inference: initial conditions and the cosmic web}

\author[Florent Leclercq \& Benjamin Wandelt]   
{Florent Leclercq$^{1,2,3}$ \and Benjamin Wandelt$^{1,2,4}$}

\affiliation{$^{1}$ Institut d'Astrophysique de Paris (IAP), UMR 7095, CNRS - UPMC Universit\'e Paris 6,\\
 98bis boulevard Arago, F-75014 Paris, France
\\[\affilskip] $^{2}$ Institut Lagrange de Paris (ILP), Sorbonne Universit\'es,\\
98bis boulevard Arago, F-75014 Paris, France
\\[\affilskip]$^{3}$ \'Ecole polytechnique ParisTech,\\
Route de Saclay, F-91128 Palaiseau, France
\\[\affilskip]$^{4}$ Departments of Physics and Astronomy,\\
University of Illinois at Urbana-Champaign, Urbana, IL 61801, USA
\\[\affilskip]emails: {\tt florent.leclercq@polytechnique.org}, {\tt wandelt@iap.fr}}

\pubyear{2014}
\volume{306}  
\pagerange{119--126}
\jname{Statistical Challenges in 21st Century Cosmology}
\editors{A.F. Heavens, J.-L. Starck \& A. Krone-Martins, eds.}
\begin{document}

\maketitle

\begin{abstract}
We describe an innovative statistical approach for the \textit{ab initio} simultaneous analysis of the formation history and morphology of the large-scale structure of the inhomogeneous Universe. Our algorithm explores the joint posterior distribution of the many millions of parameters involved via efficient Hamiltonian Markov Chain Monte Carlo sampling. We describe its application to the Sloan Digital Sky Survey data release 7 and an additional non-linear filtering step. We illustrate the use of our findings for cosmic web analysis: identification of structures via tidal shear analysis and inference of dark matter voids.
\end{abstract}

\firstsection 
\section{Introduction}

How did the Universe begin? How do we understand the shape of the present-day cosmic web? Within standard cosmology, we have an observationally well-supported model for the initial conditions (ICs) -- a Gaussian random field -- and the evolution and growth of cosmic structures is well-understood in principle. It is therefore natural to analyze large-scale structure (LSS) surveys in terms of the simultaneous constraints they place on the statistical properties of the initial conditions of the Universe and on the shape of the cosmic web. Due to the computational challenge and to the lack of detailed physical understanding of the non-Gaussian and non-linear processes that link galaxy formation to the large-scale dark matter distribution, the current state of the art of statistical analyses of LSS surveys is far from this ideal and these problems are addressed in isolation. Here, we describe progress towards the full reconstruction of four-dimensional states and illustrate the use of these results for cosmic web classification.

\section{Statistical approach}

\subsection{Why Bayesian inference?}

Cosmological observations are subject to a variety of intrinsic and experimental uncertainties (incomplete observations~-- survey geometry and selection effects~--, cosmic variance, noise, biases, systematic effects), which make the inference of signals a fundamentally ill-posed problem. For this reason, no unique recovery of the initial conditions and of the shape of the present-day cosmic web is possible; it is more relevant to quantify a probability distribution for such signals, given the observations. Adopting this point of view for large-scale structure surveys, Bayesian forward modeling (gravitational structure formation is the generative model for the complex final state, starting from a simple initial state -- Gaussian or nearly-Gaussian ICs) offers a conceptual basis for dealing with the problem of inference in presence of uncertainty \citep[e.g.][]{JascheWandelt2013,Kitaura2013,Wang2013}.

\subsection{High-dimensionality}

Statistical analysis of LSS surveys requires to go from the few parameters describing the homogeneous Universe to a point-by-point characterization of the inhomogeneous Universe. The latter description typically involves tens of millions of parameters: the density in each voxel of the survey volume. No obvious reduction of the problem size exists. ``Curse of dimensionality" phenomena \citep{Bellman1961} are therefore the significant obstacle in this high-dimensional data analysis problem. They refer to the problems caused by the exponential increase in volume associated with adding extra dimensions to a mathematical space, and therefore in sparsity given a fixed amount of sampling points. Numerical representations of high-dimensional probability distribution functions (pdfs) will tend to have very peaked features and narrow support, which means that traditional sampling methods will fail. However, gradients of these functions carry capital information, as they indicate the direction to high-density regions, permitting fast travel through a very large volume in parameter space.

\subsection{Hamiltonian Monte Carlo}

The Hamiltonian Monte Carlo algorithm \citep{Duane1987} is an algorithm for exploring parameter spaces with particles (samples). The general idea is to use classical mechanics to solve statistical problems. The algorithm interprets the negative logarithm of the pdf to sample from, $\mathcal{P}(\textbf{x})$, as a potential, $\psi \equiv - \ln(\mathcal{P}(\textbf{x}))$, and integrates Hamilton's equation in parameter space. Due to the conservation of energy in classical mechanics, the theoretical acceptance rate is always unity. Therefore, HMC beats the ``curse of dimensionality" by exploiting gradients ($\partial \psi(\textbf{x})/\partial \textbf{x}$ in Hamilton's equations) and using conserved quantities.

\section{Physical reconstructions}

\subsection{Bayesian large-scale structure inference in the SDSS DR7}
\label{sec:borg_sdss}

The full-scale Bayesian inference code \textsc{borg} (Bayesian Origin Reconstruction from Galaxies, \citealt{JascheWandelt2013}) uses HMC for four-dimensional inference of density fields in the linear and mildly non-linear regime. The (approximate) physical model for gravitational dynamics included in the likelihood is second-order Lagrangian perturbation theory (2LPT), linking initial density fields (at a scale factor $a=10^{-3}$) to the presently observed large-scale structure (at $a=1$). The galaxy distribution is modeled as a Poisson sample from these evolved density fields. The algorithm also accounts for luminosity dependent galaxy biases \citep{JascheWandelt2013b}. In \cite{JLW2014}, we apply the \textsc{borg} code to 463,230 galaxies from the \texttt{Sample dr72} of the New York University Value Added Catalogue (NYU-VAGC, \citealt{NYUVAGC}), based ot the final data release (DR7) of the Sloan Digital Sky Survey (SDSS, \citealt{SDSS,SDSS2}).

\begin{figure}
\begin{center}
\includegraphics[width=\textwidth]{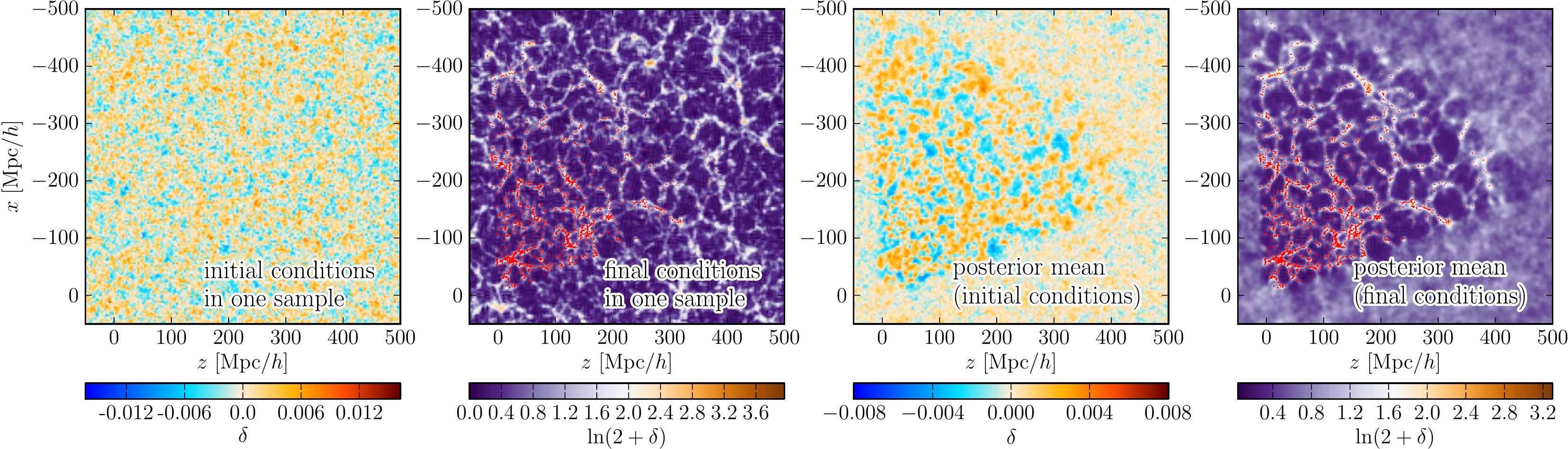}
\end{center}
\caption{Bayesian LSS inference with \textsc{borg} in the SDSS DR7. Slices through one sample of the posterior for the initial and final density fields (left) and posterior mean in the initial and final conditions (right). The input galaxies are overplotted on the final conditions as red dots.}
\label{fig:borg}
\end{figure}

Each inferred sample (Fig. \ref{fig:borg}, left) is a ``possible version of the truth" in the form of a full physical realization of dark matter particles. The variation between samples (Fig. \ref{fig:borg}, right) quantifies joint and correlated uncertainties inherent to any cosmological observation and accounts for all non-linearities and non-Gaussianities involved.

\subsection{Non-linear filtering}
\label{sec:filter}

\begin{figure}
\begin{center}
\includegraphics[width=0.6\textwidth]{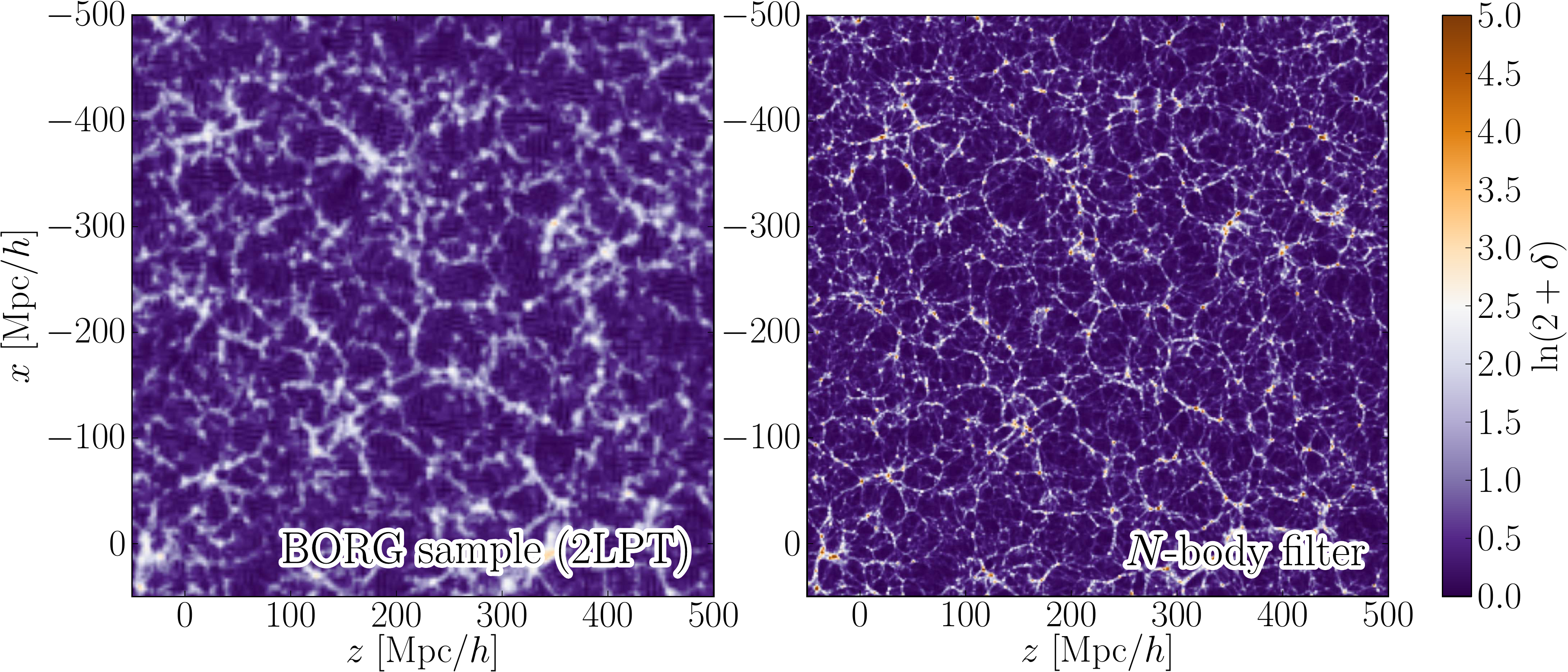} 
\end{center}
\caption{$N$-body filtering of a \textsc{borg} sample (left), to produce a non-linear data-constrained realization of the redshift-zero large-scale structure (right).}
\label{fig:filter}
\end{figure}

Building upon these results, it is possible to post-process the samples using fully non-linear dynamics as an additional filtering step \citep{LJSHW2014}. We generate a set of data-constrained realizations of the present large-scale structure: some samples of inferred initial conditions are evolved with 2LPT to $z=69$, then with a fully non-linear cosmological simulation (using \textsc{gadget-2}) from $z=69$ to $z=0$. This filtering step yields a much more precise view of the deeply non-linear regime of cosmic structure formation, sharpening overdense, virialized structures and resolving more finely the substructure of voids (Fig. \ref{fig:filter}).

\section{Cosmic web analysis}

\subsection{Tidal shear classification}

The results presented in $\S$ \ref{sec:borg_sdss} form the basis of the analysis of \cite{LJW2014}, where we classify the cosmic large scale structure into four distinct web-types (voids, sheets, filaments and clusters) and quantify corresponding uncertainties. We follow the dynamic cosmic web classification procedure proposed by \cite{Hahn2007}, based on the eigenvalues $\lambda_1 < \lambda_2 < \lambda_3$ of the tidal tensor $T_{ij}$, Hessian of the rescaled gravitational potential: $T_{ij} \equiv \partial^2 \Phi / \partial \textbf{x}_i \, \partial \textbf{x}_j$, where $\Phi$ follows the Poisson equation ($\nabla^2 \Phi = \delta$). A voxel is in a cluster (resp. in a filament, in a sheet, in a void) if three (resp. two, one, zero) of the $\lambda$s are positive.

\begin{figure}
\begin{center}
\includegraphics[width=\textwidth]{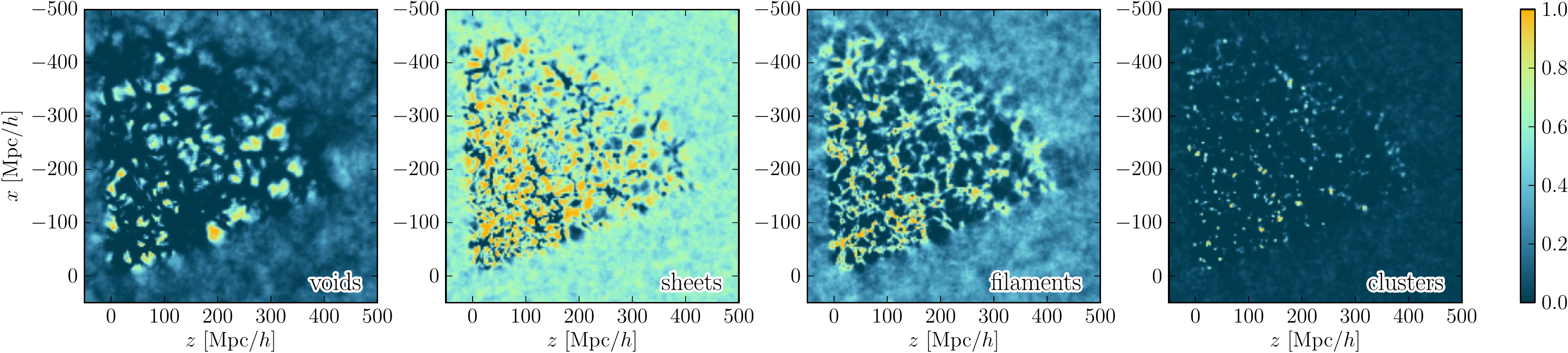}
\end{center}
\caption{Mean of the posterior pdf for the four different web-types in the SDSS DR7.}
\label{fig:ts}
\end{figure}

By applying this classification procedure to all density samples, we are able to estimate the posterior of the four different web-types, conditional on the observations. The means of these pdfs are represented in Fig. \ref{fig:ts}.

\subsection{Dark matter voids}

In \cite{LJSHW2014}, we apply computational geometry tools (\textsc{vide}: the Void IDentification and Examination pipeline, \citealt{Sutter2014}) to the constrained parts of the non-linear realizations described in $\S$ \ref{sec:filter}. We find physical cosmic voids in the field traced by the dark matter particles, probing a level deeper in the mass distribution hierarchy than galaxies. Due to the high density of tracers, we find about an order of magnitude more voids at all scales than the voids directly traced by the SDSS galaxies. In this fashion, we circumvent the issues due to the conjugate and intricate effects of sparsity and biasing on galaxy void catalogs \citep{Sutter2013} and drastically reduce the statistical uncertainty. For usual void statistics (number count, radial density profiles, ellipticities), all the results we obtain are consistent with $N$-body simulations prepared with the same setup.

\begin{acknowledgments}
We thank Jacopo Chevallard, Jens Jasche, Nico Hamaus, Guilhem Lavaux, Emilio Romano-D\'iaz, Paul Sutter and Alice Pisani for a fruitful collaboration on the projects presented here. FL acknowledges funding from an AMX grant (\'Ecole polytechnique) and BW from a senior Excellence Chair by the Agence Nationale de la Recherche (ANR-10-CEXC-004-01). This work made in the ILP LABEX (ANR-10-LABX-63) was supported by French state funds managed by the ANR within the Investissements d'Avenir programme (ANR-11-IDEX-0004-02).
\end{acknowledgments}

\end{document}